\documentclass[times,12pt]{article}
\pdfoutput=1
\usepackage{amsmath,amssymb,amsfonts,latexsym,amsthm,enumerate,url}
\usepackage{booktabs,subcaption,dcolumn}
\usepackage{graphicx}
\usepackage{mathrsfs}
\usepackage{braket}
\usepackage{xcolor}
\usepackage{float}
\usepackage{hyperref}
\usepackage{mathtools}
\usepackage{adjustbox}
\usepackage{graphicx}
\date{}

\newtheorem{thm}{Theorem}[section]

\newtheorem{lem}[thm]{Lemma}

\newcommand{\beq}[1]{\begin{equation}\label{#1}}
\newcommand{\enq}[0]{\end{equation}}

\newcommand{\remove}[1]{}

\newcommand{\comment}[1]{}

\title{Analysis of asymmetric errors in NISQ experiments\footnote {Research supported by ERC grant 834735.}}
\author{Gil Kalai, Tomer Shoham, and Carsten Voelkmann}

\begin{document}
\maketitle

\begin{abstract}
This paper continues our analysis of NISQ experiments and especially the Google 2019 ``quantum supremacy" experiment: 
In \cite {RSK22} we proposed statistical tools to analyze symmetric and asymmetric readout errors, and the symmetric case was extensively developed and studied in \cite {KRS24}.
Here we extend and complement our study from \cite {RSK22,KRS24} and analyze asymmetric readout errors for data from experiments and simulations.
Our study was further motivated by a paper by Fefferman et al. \cite {FGG+23}, who showed that the effect of amplitude damping gate errors is to have larger fractions of zeroes in the samples produced by the quantum computer.
We show that this effect of gate errors is not observed in the Google experimental data, where, for the entire range of the number of qubits, $12 \le n \le 53$, the fraction of zeroes and ones reflects only the asymmetry of reported readout errors, and report on simulations with $n=12,14$ using IBM's and Google's simulators.  

We extend our Fourier tools from \cite{KRS24} to the study of asymmetric errors. 
We also provide some analysis on the asymmetry and stability of fractions of zeroes and ones in experimental data from a few NISQ experiments.

\end {abstract}

\section {Introduction: Estimating parameters for asymmetric errors }
\label {s:asymm}

\subsection {Asymmetric readout errors and a refined noise model}

Consider a quantum circuit $C$ with $n$ qubits and let ${\cal P}_C(x)$ be the probability distribution on 0-1 bitstrings described by $C$. 
Consider a noise operation $R(q_{1\rightarrow 0}, q_{0\rightarrow 1})$ where, with probability $q_{1\rightarrow 0}$ '1' is replaced by '0' and with probability $q_{0\rightarrow 1}$ '0' is replaced by '1'.
We will denote $q=(q_{1\rightarrow 0}+q_{0\rightarrow 1})/2$, and $M=2^n$.
Following our previous papers \cite {RSK22} (Section 6) and \cite {KRS24} we will consider a 3-parameter noise model of the form

\begin {equation} 
\label {e:asy}
R(q_{1\rightarrow 0}, q_{0\rightarrow 1})(s{\cal P}_C + (1-s)/M).
\end {equation}

When we assume that $q_{1\rightarrow 0}=q_{0\rightarrow 1}=q$, our principal model reduces to the 2-parameter model:
\begin {equation}
\label {e:symm}
sR(q,q) {\cal P}_C + (1-s)/M. 
\end {equation}

For both these noise models, we developed in \cite {RSK22} estimators (primarily, maximum likelihood estimators, MLE) for the parameters based on samples from the quantum computer, and our principal example was empirical data from Google's 2019 ``quantum supremacy" experiment.
In \cite {KRS24} Fourier--Walsh expansion was used to study the 2-parameter model and to present efficient algorithms to estimate its parameters.
We referred in \cite {KRS24} to the value of $q$ that best fits the data for the symmetric noise model as the ``effective readout error".
We also studied in \cite {KRS24} simulations by Google's and IBM's quantum simulators,
data from 5-qubit circuits that we ran on IBM's 7-qubit computers ``Nairobi" and ``Jakarta",
and data from the 2023 Harvard/QuEra neutral atom experiment \cite {Blu+23}. 

One finding from \cite {KRS24} was that the effective readout error for the Google empirical data was quite close to (and, in fact, somewhat lower than) the average readout error $q=0.038$ reported in \cite {Aru+19}.
In contrast, for simulations conducted both on Google's ``QVM Weber" simulator and on IBM's ``Fake Guadalupe" simulator the effective readout error was considerably higher than the ``physical" readout error and apparently includes an additional contribution of gate errors. For the study based on Google's simulator, we considered both a noise model based on depolarizing gate errors and a more realistic noise model. 
Having a larger effective readout error compared to the physical readout error could have been expected from theoretical analysis by Gao et al. \cite {Gao+21} and Aharonov et al. \cite {AGLLV22}, and (conjecturally) from the work of Fefferman et al. \cite {FGG+23} that we discuss in section \ref {s:Fefferman}.  In this paper we study the 3-parameter model and part of our motivation came from a recent paper \cite {FGG+23} by Fefferman et al. that asserts that 
amplitude damping gate errors will have a similar effect to asymmetric 
readout errors where 1 is read as 0. 

\subsection{Fefferman  et al.'s paper on non-unital gate errors}
\label {s:Fefferman}

Fefferman, Ghosh, Gullans, Kuroiwa, and Sharma showed in \cite {FGG+23} that the effect of amplitude damping gate errors (basically the $T_1$ decay due to gates) is to have larger fractions of zeroes in the samples produced by the quantum computer.
This finding suggested to further study the asymmetric 3-parameter model for the Google experimental data, for simulations, and for data coming from other NISQ experiments.
It also suggested further potential theoretical support that the effective readout error will be larger than the physical readout error.

A central problem studied in \cite {FGG+23} is:
``Do random quantum circuits, under the influence of physically motivated noise models, anticoncentrate?" 

Here, anticoncentration means that the distribution is not concentrated on a sufficiently small number of outcomes, and it is an important property for random quantum circuit samples (RCS) under basic noise models.
In mathematical terms, if the probability for a bitstring $x$ is denoted by $p(x)$, anticoncentration refers to a situation where 
$$\displaystyle \sum_x p(x)^2 = c \frac {1}{2^n},$$ for some $c \ge 1$ is bounded as a function of $n$.
When the probability distribution is uniform, we have $c=1$, and for Porter--Thomas distributions we have $c=2$. 
Fefferman et al. showed that non-unital gate errors lead to lack of anticoncentration, namely, to large values of $c$ that diverges with the number of qubits $n$. The paper specifically studies the amplitude damping noise that is closely related to a $T_1$ gate decay.

We note that when  $q_{1\rightarrow 0} \ne q_{0\rightarrow 1}$,
our basic model of asymmetric readout errors \eqref {e:asy} 
exhibits, for large values of $n$, a lack of anticoncentration, and we conjecture that the effect of non-unital gate errors amounts effectively to asymmetric readout errors given by \eqref {e:asy}, with $q_{1\rightarrow 0} > q_{0\rightarrow 1}$.

{\bf Remarks:} Our study of the Google data shows that the value of $c$ (which is estimated by $1+T^2$, where $T$ is our estimator from Section 4.7 of \cite {RSK22}), is substantially larger than what is expected from the Google noise model.
(The Google noise model would give $c=1+\phi^2$, where $\phi$ is the fidelity.)
This effect appears to be stronger than the effect of asymmetric readout errors and it is not caused by additional bias towards 0.
(While $T^2/\phi^2$ increases rather rapidly, the value of $T^2$ itself still decreases as a function of $n$.)

We note that the authors of \cite {FGG+23} expected ``lack of anticoncentration" in a regime where the error rate is higher than $1/n$ but not when the error rate is considerably smaller than $1/n$.
(A similar distinction goes back to Kalai and Kindler \cite {KalKin14}.)
A more delicate question is whether the effect of non-unital gate errors can be detected for smaller values of $n$ even if the effect on $c$ is rather small.

\subsection {Some research problems} 
\label {s:prob}

Our central research problem is to what extent the gate errors in samples coming from quantum computers lead especially to higher values of the estimated parameter $q_{1\rightarrow 0}$ and thus express the findings of Fefferman et al. \cite {FGG+23}. We now list other related research problems.  

\begin {enumerate} \item 
Does the refined MLE estimation for the parameters $(s, q_{1\rightarrow 0}, q_{0\rightarrow 1}$) in the 3-parameter noise model agree with the MLE estimation of the parameters $(s,q)$ based on the 2-parameter noise model?

\end {enumerate}

\begin {enumerate} 
\item[2.]
How to apply Fourier--Walsh methods to express the noise model described by Eq.~\eqref {e:asy}, and to efficiently estimate the parameters $s, q_{1\rightarrow 0}, q_{0\rightarrow 1}$?
\end {enumerate}

\begin {enumerate}
\item[3.]
What do we learn from estimating these parameters for Google's experimental data, for data from other NISQ experiments, and for simulations?

\item [4.] The analysis from \cite {KRS24}, both for the Google data and the Google QVM simulator, depends on the average readout errors reported in \cite {Aru+19}. Do the conclusions hold when we replace the average values with the detailed reported individual values?
  
\end {enumerate}

\subsection {Our findings} 

In Section \ref {s:google} we study the bias towards zeroes for the Google experimental data and from data from other quantum computers.
The effect of gate errors toward larger ratios of zeroes is not witnessed in the Google experimental data, for the entire range of 12-53 qubits.

We also report on our experimental data for IBM 5-qubit quantum computers. 
In Section \ref {s:google-sim} we consider data from Google and IBM quantum simulators. The Google team uploaded data leading to two sets of values for the readout errors of individual qubits. The ``effective" readout errors are similar but somewhat smaller for one set of readout errors and considerably smaller for the second. 
For the samples from the IBM simulator ``Fake Guadalupe" (12 qubits) we observed in \cite {KRS24} effective readout errors that are considerably larger than the physical readout errors. The asymmetry between the ratios of ones and zeroes is considerably larger than the difference between the reported value of $q_{0\rightarrow 1} -q_{1\rightarrow 0}$, and this may be related to the effect of gates. Because of some technical difficulties, the situation for the Google simulations is inconclusive and further simulations are required for getting a clear picture. 
Appendices \ref {s:A} and \ref {s:B} include some tables and analysis of the readout errors and effective readout errors from experiments and simulations. In Section \ref {s:fourier} of the Appendix we discuss theoretical aspects of extending the Fourier framework from Ref.~\cite {KRS24} to the case of asymmetric readout errors. 
In Section \ref {s:temporal_instability} we study temporal variation of the proportion of measured 1’s during the measurement of bitstrings.

\section {Analyzing data from quantum computers and simulators}
\label {s:google}

One basic form of errors are readout errors: the probability that a measured qubit will read as '0' instead of '1' and vice versa. For the Google 2019 experiment, the supplement and supplementary data of \cite {Aru+19} report that the average probability $ q_{1\rightarrow 0}$ that 1 is read as 0 is 0.055 and the average probability $ q_{0\rightarrow 1}$ that 0 is read as 1 is 0.023.

\subsection {Estimating $q_{1\rightarrow 0}$ and $q_{0\rightarrow 1}$}
\label {s:est-asym}
There are several ways to estimate the values of $q_{1\rightarrow 0}$ and $q_{0\rightarrow 1}$.

\begin {enumerate}
\item 
We can get an estimate on $q_{1\rightarrow 0}-q_{0\rightarrow 1}$ based on the statistics of 0's and 1's in the empirical samples.
Namely, the expected number of 0's is $1/2+(q_{1\rightarrow 0}-q_{0\rightarrow 1})/2$.
(This gives estimates for individual qubits as well.)

\item
The readout error rates were reported in the file \texttt{som\_params\_by\_qubit.csv} (uploaded in January 2020) in Google's supplementary data to \cite {Aru+19}.
The values of the readout errors from this file are given in Tables \ref {t:google-readout-data} and \ref {t:google-readout-data-av}. The Google team uploaded (in January 2021) further raw data (file \texttt{readout\_raw\_data.tar}) for readout errors based on the initialization of computational basis states and then measuring them.
This method provides the values of the readout errors for individual qubits which are given (in parentheses) in Table \ref {t:google-readout-data-av}..

\item 
We developed a statistical method (based on knowing the amplitudes of the ideal circuit) to estimate $q_{1\rightarrow 0}$ and $q_{0\rightarrow 1}$ and implemented it for one circuit with 12 qubits (see Section~6.4 in Ref.~\cite {RSK22}).

\end {enumerate} 

Note that the first and third methods account not only for bias toward 0 for readout errors but also for additional bias toward 0 coming from non-unital gate errors that is considered in Fefferman et al.'s paper \cite {FGG+23}. It appears that Google's method (item 2) does not involve the effect of non-unital gate errors, since the qubits are initialized and then are measured.

\subsection {Are amplitude damping gate errors manifested for the Google 2019 data?}

We compared the different methods for estimating the parameters $ q_{1\rightarrow 0}$ and $ q_{0\rightarrow 1}$.
Recall that Google's estimates were $q_{1\rightarrow 0}=0.055$ and $q_{0\rightarrow 1}=0.023$.
We observed that

\begin {itemize}
\item
For $12 \le n \le 53$, the difference between the fraction of 0's and the fraction of 1's
in the Google experimental data is {\it not larger} than the effect of readout errors $ q_{1\rightarrow 0}, \, q_{0\rightarrow 1}$ as estimated in the Google data.
As a matter of fact, the empirical difference between 0's and 1's is roughly 10\%-20\% lower than $q_{1\rightarrow 0}-q_{0\rightarrow 1}$.
See Table \ref {t:google-readout-data-av}.
No additional effect from gate errors to a larger fraction of 0's is observed for the entire range $12 \le n \le 53$.
\item 
For $n=12,14$ the MLE estimations for $q_{1\rightarrow 0}$ and $q_{0\rightarrow 1}$ are similar (or slightly lower) to the average physical readout errors as estimated by Google's method. Again, no additional effect from gate errors is observed.  

For $n=12$ the Google readout data gives $q_{1\rightarrow 0}=0.050$ and $q_{0\rightarrow 1}=0.021$ (Table~\ref{t:google-readout-data-av})
and the averaged MLE estimated values are $q_{1\rightarrow 0}=0.046$ and $q_{0\rightarrow 1}=0.022$ (Table~\ref{t:googleexp}).
For $n=14$ the Google readout data gives $q_{1\rightarrow 0}=0.046$ and $q_{0\rightarrow 1}=0.022$ (Table~\ref{t:google-readout-data-av})
and the averaged MLE estimated values are $q_{1\rightarrow 0}=0.042$ and $q_{0\rightarrow 1}=0.020$ (Table~\ref{t:googleexp}).
These MLE values for $q_{1\rightarrow 0}$ and $q_{0\rightarrow 1}$ are consistent with the estimated value of $q$ for the symmetric model.
\end {itemize}

Both these findings show no sign in the Google experimental samples for additional asymmetry between 0's and 1's caused by amplitude damping gate errors (or any other cause).

\subsection {Data from other quantum computer experiments}

For our 5-qubit experiments on Nairobi and Jakarta, the average physical readout errors were 0.0256 and 0.0247, respectively;
the average probabilities for '1' to be read as '0' were (0.0373, 0.0360) and for '0' to be read as '1' were (0.0138, 0.0135).
(These values are based on the five qubits of our quantum circuits.) 
Our MLE computation of effective readout errors gave for Nairobi the value 0.044 and for Jakarta the value 0.067 (see Table~\ref {t:ibm_q_nj}). The finer MLE asymmetric estimates gave for Nairobi $q_{1\rightarrow 0}=0.077$ and $q_{0\rightarrow 1}=0.027$, and for Jakarta $q_{1\rightarrow 0}=0.075$ and $q_{0\rightarrow 1}=0.044$.

\subsection {Data from simulators}
\label{s:google-sim}

For IBM's Fake Guadalupe simulator, the effective readout error estimations are given in Table \ref{table:fake_s_q}.
While the average physical readout error (for the 12 qubits used for the circuit) was 0.0218,
the average MLE estimated value for $n=12$ was $q=0.066$.
The finer MLE estimates for the errors from 1 to 0 and from 0 to 1, based on the same data give $q_{1\rightarrow 0}=0.081$ and $q_{0\rightarrow 1}=0.048$ compared to the physical values of 0.0278 and 0.0072, respectively.
The average value of $q_{1\rightarrow 0}- q_{0\rightarrow 1}$ is 0.0207 for the reported readout errors and 0.033 for the MLE estimations. The gap between the ratios of ones and zeroes is even larger --- 0.042.
For the Google QVM Weber simulator the MLE estimates give for $n=12$, $q_{1\rightarrow 0}=0.077$ and $q_{0\rightarrow 1}=0.020$ and for $n=14$, $q_{1\rightarrow 0}=0.076$ and $q_{0\rightarrow 1}=0.020$, see Table \ref {t:weber}. However, due to some ambiguity regarding the values of errors for these simulations, 
further careful simulations (also for larger circuits) are required.

\section {Conclusion}

The effect discovered by Fefferman et al. \cite {FGG+23} of amplitude damping noise on gates toward larger fraction of zeroes in the samples is not witnessed in the samples from the Google experimental data of the 2019 supremacy experiment \cite {Aru+19} for the entire range of numbers of qubits between 12 and 53.
This effect is also not observed in data from Google's Sycamore-67, -69, -70 and for USTC’s Zuchongzhi-56.
The Fefferman et al. effect appears to be witnessed in simulations on the ``Fake Guadalupe" IBM simulator for 12 qubits. For better understanding, additional samples from quantum computers and from simulations are required.

\subsection *{Acknowledgements}
Research supported by ERC grant 834735. We are thankful to our team-member Ohad Lev for running simulations on the IBM and Google simulators, running 5-qubit experiments on IBM quantum computers, and for many helpful conversations.
We thank Bill Fefferman and Soumik Ghosh for helpful discussions.
We thank Yosi Rinott 
for helpful corrections and thoughtful suggestions.

\newpage
\appendix
\section{Readout error rates} 
\label {s:A}
In this section we describe the reported readout errors from the quantum computers and simulations that we studied.
The tables from Section \ref {s:A1} are reproduced from \cite {KSV25}. 

\subsection {Google's Sycamore}
\label {s:A1}
\subsubsection*{Readout error rate reported by Google}

\begin{table}[H]
\begin{subtable}[t]{0.31\textwidth}
\resizebox{\textwidth}{!}{\begin{tabular}{||c c c c||} 
\hline
qubit & $n_{\text{ins}}$ & $q_{0\rightarrow 1}$ & $q_{1 \rightarrow 0}$\\ [0.5ex] 
\hline \hline
q0\_5&38&0.005&0.028\\
q0\_6&40&0.005&0.031\\
q1\_4&36&0.030&0.035\\
q1\_5&28&0.008&0.030\\
q1\_6&28&0.011&0.066\\
q1\_7&30&0.008&0.036\\
q2\_4&14&0.014&0.025\\
q2\_5&14&0.033&0.028\\
q2\_6&18&0.020&0.058\\
q2\_7&24&0.010&0.037\\
q2\_8&34&0.005&0.041\\
q3\_2&36&0.020&0.056\\
q3\_3&12&0.038&0.051\\
q3\_4&12&0.023&0.061\\
q3\_5&12&0.007&0.075\\
q3\_6&12&0.004&0.042\\
q3\_7&20&0.013&0.039\\
q3\_8&34&0.010&0.038\\
\hline
\end{tabular}}
    \end{subtable}   
\begin{subtable}[t]{0.31\textwidth}
\resizebox{\textwidth}{!}{\begin{tabular}{||c c c c||} 
\hline
qubit & $n_{\text{ins}}$ & $q_{0\rightarrow 1}$&$q_{1 \rightarrow 0}$\\ [0.5ex] 
\hline \hline
q3\_9&46&0.009&0.051\\
q4\_1&38&0.019&0.051\\
q4\_2&22&0.011&0.036\\
q4\_3&12&0.019&0.059\\
q4\_4&12&0.032&0.063\\
q4\_5&12&0.053&0.156\\
q4\_6&12&0.004&0.105\\
q4\_7&20&0.015&0.078\\
q4\_8&44&0.009&0.060\\
q4\_9&50&0.015&0.061\\
q5\_0&53&0.013&0.056\\
q5\_1&32&0.019&0.054\\
q5\_2&22&0.012&0.094\\
q5\_3&12&0.024&0.059\\
q5\_4&12&0.018&0.051\\
q5\_5&12&0.050&0.068\\
q5\_6&12&0.130&0.075\\
q5\_7&42&0.014&0.043\\
\hline    
\end{tabular}}
\end{subtable}
\begin{subtable}[t]{0.31\textwidth}
\resizebox{\textwidth}{!}{\begin{tabular}{||c c c c||} 
\hline
qubit & $n_{\text{ins}}$ & $q_{0\rightarrow 1}$ & $q_{1 \rightarrow 0}$\\ [0.5ex] 
\hline \hline
q5\_8&51&0.013&0.046\\
q6\_1&32&0.029&0.070\\
q6\_2&24&0.031&0.033\\
q6\_3&18&0.011&0.056\\
q6\_4&16&0.039&0.043\\
q6\_5&16&0.013&0.037\\
q6\_6&39&0.092&0.046\\
q6\_7&48&0.018&0.037\\
q7\_2&30&0.018&0.042\\
q7\_3&26&0.030&0.041\\
q7\_4&26&0.030&0.048\\
q7\_5&43&0.016&0.062\\
q7\_6&47&0.031&0.092\\
q8\_3&41&0.033&0.075\\
q8\_4&45&0.020&0.059\\
q8\_5&49&0.027&0.071\\
q9\_4&53&0.015&0.049\\
\textbf{Avg}&&\textbf{0.023}&\textbf{0.055}\\
\textbf{Std}&&\textbf{0.021}&\textbf{0.023}\\
\hline
\end{tabular}}
    \end{subtable} \caption{Sycamore's asymmetric readout error rates per qubit for simultaneous operation (source: Google's file \texttt{som\_params\_by\_qubit.csv} in agreement with the lower two panels of Fig. S24 in the supplement of \cite {Aru+19}).
The column $n_{\text{ins}}$ is the number $n$ of qubits, for which the respective qubit is added to the circuit
(source: Python list QUBIT\_ORDER in the circuit description files of the full circuits).
The average for $n=12$ is $q_{1\rightarrow 0}=0.0718$, $q_{0\rightarrow 1}=0.0335$,
and for $n=14$ the averages are $q_{1\rightarrow 0}=0.0653$, $q_{0\rightarrow 1}=0.0321$ (see Table~\ref {t:google-readout-data-av}).}
\label{t:google-readout-data}
\end{table}

\newpage
\subsubsection*{Average readout error rates for Google's Sycamore}

\begin{table}[H]
\begin{center} \resizebox{0.82\textwidth}{!}{
    \begin{tabular}{||c | c c c c | c||} 
 \hline
  $n$&\multicolumn{4}{|c|}{Average readout error in \%}&Empirical \\
  &$q_{0\rightarrow 1}$&$q_{1 \rightarrow 0}$ & Avg & Diff & Diff \\ [0.5ex] 
\hline \hline
12&3.35 (2.12)&7.18 (5.04)&5.27 (3.58)&3.83 (2.92)&2.51\\
14&3.21 (2.17)&6.53 (4.60)&4.87 (3.39)&3.32 (2.43)&2.26\\
16&3.13 (2.08)&6.22 (4.55)&4.68 (3.32)&3.09 (2.47)&2.24\\
18&2.96 (2.04)&6.16 (4.97)&4.56 (3.51)&3.20 (2.93)&2.45\\
20&2.80 (2.27)&6.13 (4.59)&4.47 (3.43)&3.33 (2.32)&2.06\\
22&2.65 (2.16)&6.16 (5.00)&4.41 (3.58)&3.51 (2.84)&2.51\\
24&2.60 (2.19)&5.94 (5.14)&4.27 (3.67)&3.34 (2.95)&2.68\\
26&2.63 (2.34)&5.83 (5.24)&4.23 (3.79)&3.20 (2.90)&2.51\\
28&2.51 (2.05)&5.75 (5.17)&4.13 (3.61)&3.24 (3.12)&2.57\\
30&2.43 (2.20)&5.63 (5.15)&4.03 (3.68)&3.20 (2.95)&2.57\\
32&2.42 (2.20)&5.67 (5.07)&4.05 (3.64)&3.25 (2.87)&2.58\\
34&2.33 (2.32)&5.56 (5.74)&3.95 (4.03)&3.23 (3.42)&2.41\\
36&2.33 (2.20)&5.51 (5.15)&3.92 (3.68)&3.18 (2.95)&2.66\\
38&2.27 (2.18)&5.42 (5.40)&3.85 (3.79)&3.15 (3.22)&2.89\\
39&2.45 (2.16)&5.40 (5.53)&3.93 (3.85)&2.95 (3.37)&2.78\\
40&2.40 (2.16)&5.34 (5.66)&3.87 (3.91)&2.94 (3.50)&3.03\\
41&2.43 (2.27)&5.40 (5.51)&3.92 (3.89)&2.97 (3.24)&2.98\\
42&2.40 (2.36)&5.37 (5.47)&3.89 (3.92)&2.97 (3.11)&2.68\\
43&2.38 (2.35)&5.39 (5.50)&3.89 (3.93)&3.01 (3.15)&3.17\\
44&2.35 (2.26)&5.40 (5.53)&3.88 (3.90)&3.05 (3.27)&2.89\\
45&2.34 (2.40)&5.41 (5.59)&3.88 (4.00)&3.07 (3.19)&3.00\\
46&2.31 (2.43)&5.41 (5.58)&3.86 (4.01)&3.10 (3.15)&2.80\\
47&2.33 (2.56)&5.49 (5.75)&3.91 (4.16)&3.16 (3.19)&2.87\\
48&2.32 (2.74)&5.45 (5.76)&3.89 (4.25)&3.13 (3.02)&2.83\\
49&2.32 (2.63)&5.49 (5.80)&3.91 (4.22)&3.17 (3.17)&2.91\\
50&2.31 (2.73)&5.50 (5.58)&3.91 (4.16)&3.19 (2.85)&2.74\\
51&2.29 (2.61)&5.48 (5.55)&3.89 (4.08)&3.19 (2.94)&2.81\\
53&2.25 (2.33)&5.47 (5.70)&3.86 (4.02)&3.22 (3.37)&2.93\\
\hline
\end{tabular}}
\caption{Columns $q_{0\rightarrow1}$ and $q_{1\rightarrow0}$ are the average readout error rates based on Table~\ref {t:google-readout-data}, each average includes only the qubits in this circuit.
In parenthesis are the readout error relative frequencies in Google's initialization/measurement file \texttt{readout\_raw\_data.tar} uploaded by Google in January 2021.
The last column gives the difference between percentage of zeroes and ones in the Google experimental samples, for all full circuits of pattern EFGH. 
This difference is smaller than the difference based on the readout data.
We do not witness in the Sycamore experimental data additional asymmetry predicted in \cite {FGG+23} that is based on non-unital gates.}
\label{t:google-readout-data-av}
\end{center}
\end{table}

Further remarks about Table \ref {t:google-readout-data-av}:
The raw data files provided by the Google team are based on initializing the quantum computer (with many computational basis states) and then measuring.
For example, for $n=12$ there were 108,000 bitstrings arising from 36 distinct computational basis states,
and for $n=53$, there were 477,000 samples from 159 distinct states.
We suppose that readout errors provided in the supplementary data itself are based on a similar method.
We do not know the reason for the differences between the readout error parameters from the supplementary data (file \texttt{som\_params\_by\_qubit.csv}) and from the additional raw data (file \texttt{readout\_raw\_data.tar}).   

\subsection{IBM's Fake Guadalupe simulator}

\begin{table}[H]
\begin{center}
\begin{subtable}[t]{0.6\textwidth}
    \begin{tabular}{||c c c||} 
 \hline
Qubit&$q_{1\rightarrow 0}$&$q_{0 \rightarrow 1}$\\ [0.5ex] 
\hline \hline
1&0.0214&0.0058\\
2&0.0352&0.0058\\
3&0.0234&0.0092\\
4&0.0434&0.0196\\
5&0.0242&0.0054\\
6&0.0334&0.0060\\
7&0.0214&0.0016\\
8&0.0308&0.0098\\
9&0.0912&0.0298\\
\hline
\end{tabular}
\end{subtable}   \begin{subtable}[t]{0.6\textwidth}
    \begin{tabular}{||c c c||} 
 \hline
Qubit&$q_{1\rightarrow 0}$&$q_{0 \rightarrow 1}$\\ [0.5ex] 
\hline \hline
10&0.0266&0.0044\\
11&0.0250&0.0074\\
12&0.0232&0.0056\\
13&0.0334&0.0048\\
14&0.0186&0.0048\\
15&0.0374&0.0044\\
16&0.0188&0.0024\\
\textbf{Avg}&\textbf{0.0317}&\textbf{0.0079}\\
\textbf{Std}&\textbf{0.0169}&\textbf{0.0069}\\
\hline
\end{tabular}
\end{subtable}   
\caption{Readout error rates for IBM's Fake Guadalupe simulator.
We used qubits \{1, 2, 3, 4, 5, 6, 7, 8, 11, 12, 13, 14\}.
}
\label{t:re-guadalupe}
\end{center}
\end{table}

\subsection {Google's QVM Weber simulator}
\label {s:reported-weber}
\begin{figure*}[!ht]
\centering
\hspace*{-2cm}
\includegraphics[width=7.1in]{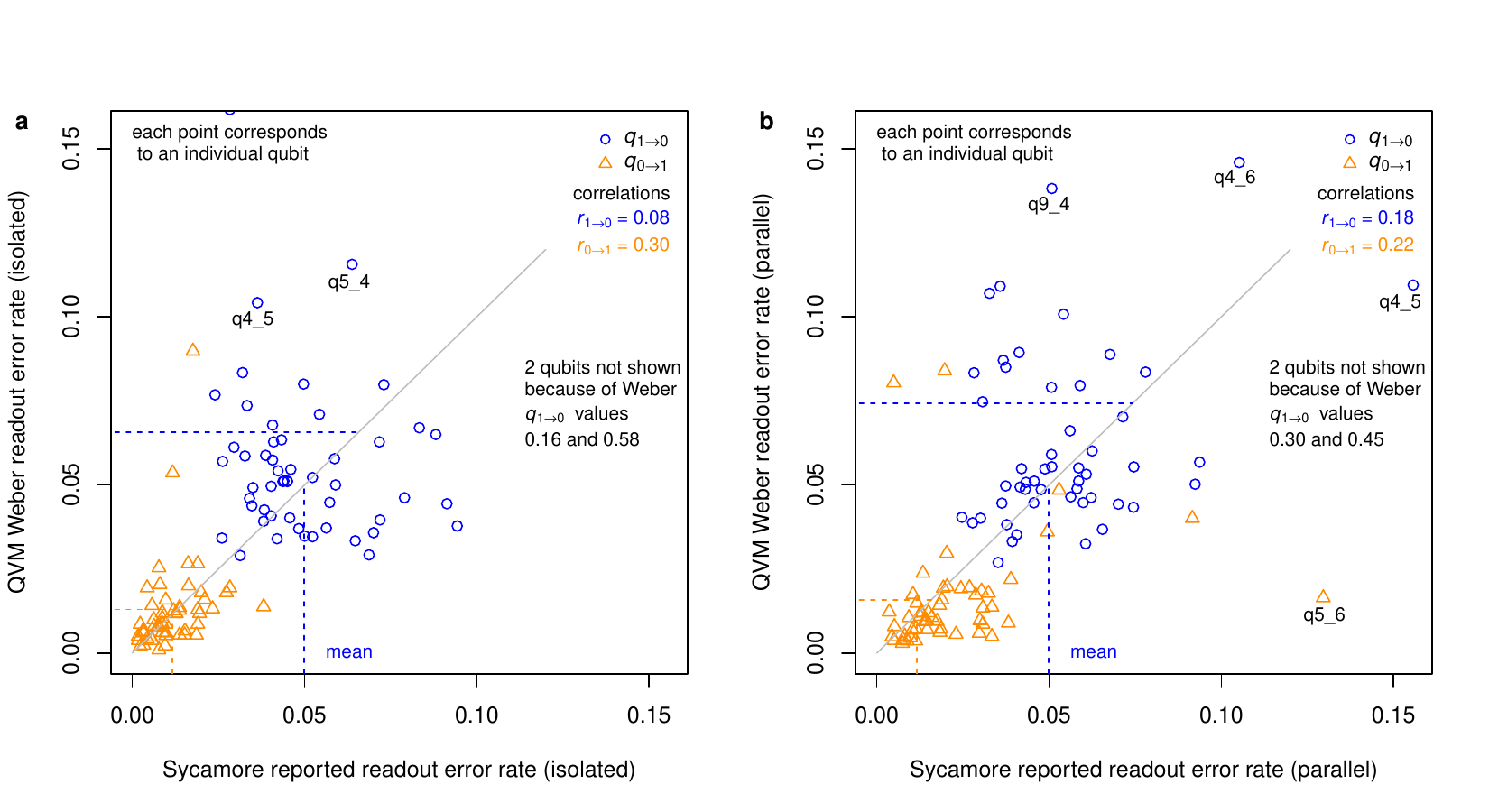}
\caption{\textbf{Asymmetric qubit readout error rates from Sycamore vs. Google's QVM Weber calibration data. a,}
Dependency between the asymmetric readout error rates reported in Google's published file \texttt{som\_params\_by\_qubit.csv} (columns \texttt{sq\_readout\_error\_\textbar0\textrangle{}} and \texttt{sq\_readout\_error\_\textbar1\textrangle})
and the readout error rates from Google's QVM Weber calibration data both obtained from isolated operation on qubits.
The source of the calibration data for Google's QVM Weber simulator is the Python command \texttt{cirq\_google.engine.load\_median\_device\_calibration("weber")} retrieved on 21 Oct 2024.
\textbf{b,}
Same as panel a, but with both Sycamore's and QVM Weber's error rates determined by parallel (not isolated) qubit operation.
Also in this case the correlations between Sycamore's reported error rates and QVM Weber's are rather low.
So the QVM Weber calibration parameters do not correspond to Sycamore's parameters.}
\label{fig_readout_err_Syc_rep_vs_Weber}
\end{figure*}

The supplement and the published data to Google's quantum supremacy paper \cite {Aru+19} as well as the calibration data of Google's QVM Weber quantum simulator both include two sets of error parameters:
one set of errors estimated with ``isolated/single" measurements one set of errors for ``simultaneous/parallel" measurements. 
Overall, the Weber asymmetric error rates of the 53 QVM Weber individual qubits have only a small correlation with the same 53 qubits of Sycamore (see Fig.~\ref{fig_readout_err_Syc_rep_vs_Weber}).

\begin{figure*}[!ht]
\centering
\includegraphics[width=4in]{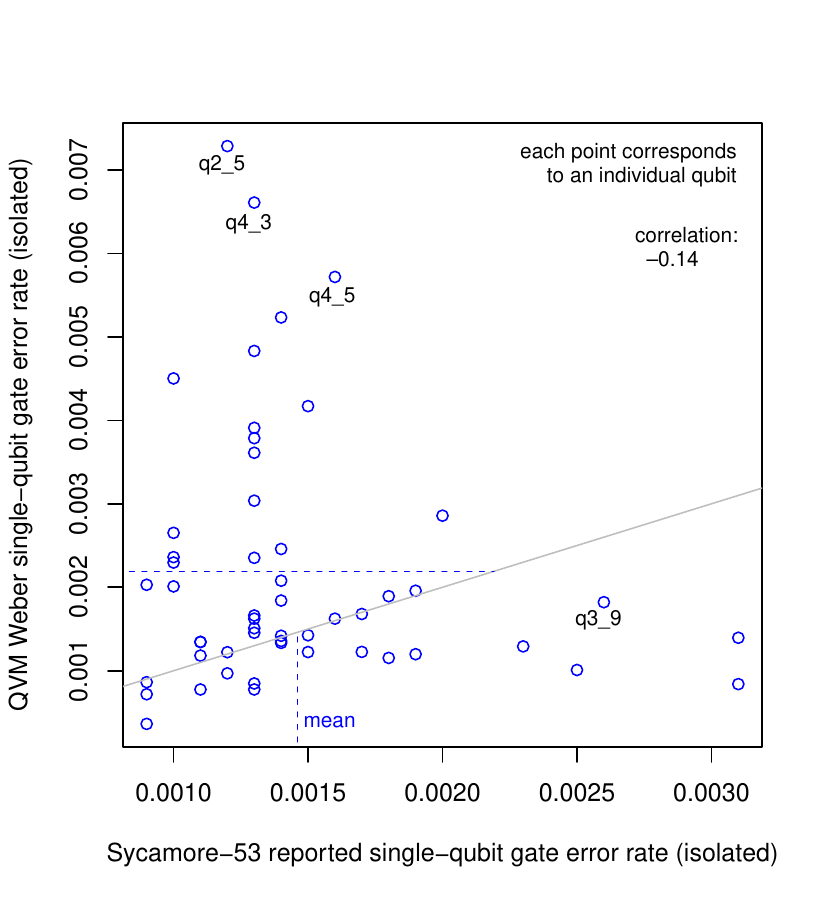}
\caption{\textbf{Single-qubit gate error rates from Sycamore-53 vs. Google's QVM Weber calibration data.}
The source of the Sycamore-53 single-qubit gate errors is Google's published file \texttt{som\_params\_by\_qubit.csv}, column \texttt{sq\_errors\_isolated}.
The source QVM Weber's single-qubit gate errors is its calibration data (Python command \texttt{cirq\_google.engine.load\_median\_device\_calibration("weber")} retrieved on 21 Oct 2024).
Obviously the QVM Weber single-qubit calibration parameters do not correspond to Sycamore's parameters.}
\label{fig_single-qubit_gate_err_Syc_rep_vs_Weber}
\end{figure*}

\clearpage

The average values of $q_{1\rightarrow 0}$ and $q_{0\rightarrow1}$ are presented in Table \ref {t:qvm-readout-data-av}.

\begin{table}[H]
\begin{center} \resizebox{1.0\textwidth}{!}{
\begin{tabular}{||c | c c c c |} 
\hline
$n$&$q_{0\rightarrow 1}$ single (parallel)&$q_{1 \rightarrow 0}$ single (parallel) & $q$ & Difference  \\ [0.5ex] 
\hline \hline
12&0.01208 (0.01550) &0.06950 (0.10046)& 0.04079 (0.05810)& 0.05742 (0.08496)\\
14&0.01143 (0.01432) &0.06750 (0.09495) &0.03946 (0.05464)& 0.05607 (0.08063)\\
53&0.01293 (0.01583) &0.06565 (0.07424) & 0.03929 (0.04504)& 0.05272 (0.05840)\\
\hline
\end{tabular}}
\caption{The second and third columns present the average $q_{1\rightarrow 0}$ and $q_{0\rightarrow 1}$ ``single" readout errors over the circuit's qubits, in parenthesis are the corresponding ``parallel" readout errors
(source of these error rates: \texttt{cirq\_google.engine.load\_median\_device\_calibration("weber"))}.
The fourth and fifth columns give the average $q$ and difference of these values. 
}
\label{t:qvm-readout-data-av}
    \end{center}
\end{table}

\begin{table}[H]
\begin{center} \resizebox{1.0\textwidth}{!}{
\begin{tabular}{||c | c c |} 
\hline
&Google's 2019 experiment & QVM Weber\\ [0.5ex] 
\hline
$q_{1 \rightarrow 0}$ isolated/single&0.050&0.066\\
$q_{1 \rightarrow 0}$ simultaneous/parallel&0.055&0.074\\
$q_{0\rightarrow 1}$ isolated/single& 0.012&0.013\\
$q_{0\rightarrow 1}$ simultaneous/parallel&0.023&0.016\\  
\hline \hline

\end{tabular}}
\caption{Asymmetric readout error rates averaged over all $n=53$ qubits for reported by Google for their 2019 experiment and for the QVM Weber simulator.
(Source for Google's 2019 experimental readout errors: Google's file \texttt{som\_params\_by\_qubit.csv}, Table II and Fig. S24 in the supplement of \cite {Aru+19}).
The average simultaneous readout error reported in 2019 (0.038, see \cite {Aru+19}) is close to the average readout error for the simulator in the ``single" category (0.039).}
\label{t:com-readout-data-av}
\end{center}
\end{table}

\subsection{Nairobi and Jakarta}

\begin{table}[H]
\begin{subtable}[t]{0.6\textwidth}
    \begin{tabular}{||c c c||} 
     \multicolumn{3}{c}{\textbf{Nairobi}} \\
     \multicolumn{3}{c}{} \\
 \hline
Qubit&$q_{1\rightarrow 0}$&$q_{0 \rightarrow 1}$\\ [0.5ex] 
\hline \hline
0&0.0328&0.0132\\
1&0.0352&0.0128\\
2&0.0536&0.0122\\
3&0.0448&0.0108\\
4&0.0334&0.0104\\
5&0.0378&0.0218\\
6&0.0360&0.0106\\
\textbf{Avg}&\textbf{0.0391}&\textbf{0.0131}\\
\textbf{Std}&\textbf{0.0075}&\textbf{0.0040}\\
\hline
\end{tabular}
    \end{subtable}   \begin{subtable}[t]{0.6\textwidth}
\begin{tabular}{||c c c||} 
     \multicolumn{3}{c}{\textbf{Jakarta}} \\
     \multicolumn{3}{c}{} \\
 \hline
Qubit&$q_{1\rightarrow 0}$&$q_{0 \rightarrow 1}$\\ [0.5ex] 
\hline \hline
0&0.0320&0.0088\\
1&0.0250&0.0068\\
2&0.0282&0.0062\\
3&0.0278&0.0084\\
4&0.0338&0.0628\\
5&0.0504&0.0306\\
6&0.0486&0.0154\\
\textbf{Avg}&\textbf{0.0351}&\textbf{0.0199}\\
\textbf{Std}&\textbf{0.0103}&\textbf{0.0208}\\
\hline    
\end{tabular}
\end{subtable}
\caption{Readout error rates for all qubits of IBM's Nairobi and Jakarta 7-qubit quantum processors.
We used the 5 qubits \{0, 1, 3, 5, 6\} for Nairobi and the 5 qubits \{1, 2, 3, 5, 6\} for Jakarta.
}
\label{t:name11}
\end{table}

\clearpage 
\section {Estimated vs. reported parameters}
\label {s:B}
\subsection{Google's Sycamore}
The reported average readout errors for the 12-qubit circits are 
for $n=12$ the reported Google readout data gives $q_{1\rightarrow 0}=0.072$ and $q_{0\rightarrow 1}=0.053$.
For $n=14$ the reported Google readout data gives $q_{1\rightarrow 0}=0.065$ and $q_{0\rightarrow 1}=0.032$.
These values are computed from the data file  \texttt{som\_params\_by\_qubit.csv} uploaded by the Google team in 2020 (they are consistent with figures in the Google paper). The Google team provided an additional file \texttt{readout\_raw\_data.tar} with raw data for readout errors that gives for $n=12$ the average values $q_{1\rightarrow 0}=0.050$ and $q_{0\rightarrow 1}=0.021$, and for $n=14$ $q_{1\rightarrow 0}=0.048$ and $q_{0\rightarrow 1}=0.022$. 

Table \ref {t:googleexp} provides MLE estimations of the (effective) readout error rates for Google's Sycamore with $n=12,14$ qubits.
The averaged estimated values based on MLE and our  three-parameter model are for $n=12$ 
$q_{1\rightarrow 0}=0.046$ and $q_{0\rightarrow 1}=0.022$ and for $n=14$ the averaged estimated values are $q_{1\rightarrow 0}=0.042$ and $q_{0\rightarrow 1}=0.020$. 
The average MLE values for $q_{1\rightarrow 0}$ and $q_{0\rightarrow 1}$ are consistent with the average estimated value of $q$ for the symmetric model.

The MLE estimations for $q_{1\rightarrow 0}$ and $q_{0\rightarrow 1}$ are considerably smaller than the averaged reported values in the 2020 Google data from \texttt{som\_params\_by\_qubit.csv}.
They are similar (and somewhat smaller) to the average reported readout errors in the file \texttt{readout\_raw\_data.tar}.
In any case, no additional effect is observed from gate errors toward higher values. 

The average MLE values for $q_{1\rightarrow 0}$ and $q_{0\rightarrow 1}$ are consistent with the average estimated value of $q$ for the symmetric model.

\enlargethispage{3cm}

\begin{table}[H]
\adjustbox{max width=\textwidth}{ \centering
 \begin{tabular}[t]{||c c c c||} 
 \hline
$n$ & $i$ & $s$ & $q$\\[0.5ex] 
\hline\hline 
12&0&0.605&0.039\\
    12&1&0.555&0.033\\
    12&2&0.517&0.030\\
    12&3&0.570&0.034\\
    12&4&0.570&0.035\\
    12&5&0.648&0.046\\
    12&6&0.569&0.036\\
    12&7&0.518&0.030\\
    12&8&0.559&0.035\\
    12&9&0.541&0.031\\
\textbf{12}&\textbf{Avg}&\textbf{0.565}&\textbf{0.035}\\
\textbf{12}&\textbf{Std}&\textbf{0.037}&\textbf{0.004}\\
\hline
\textbf{12}&\textbf{Rep1}&\textbf{--}&\textbf{0.064}\\
\textbf{12}&\textbf{Rep2}&\textbf{--}&\textbf{0.035}\\
\hline
\multicolumn{4}{c}{} \\
\hline
$n$ & $i$ & $s$ & $q$\\[0.5ex] 
\hline\hline 
    14&0&0.510&0.031\\
    14&1&0.561&0.038\\
    14&2&0.508&0.031\\
    14&3&0.506&0.030\\
    14&4&0.530&0.034\\
    14&5&0.462&0.025\\
    14&6&0.484&0.028\\
    14&7&0.490&0.028\\
    14&8&0.521&0.033\\
    14&9&0.533&0.034\\
\textbf{14}&\textbf{Avg}&\textbf{0.510}&\textbf{0.031}\\
\textbf{14}&\textbf{Std}&\textbf{0.027}&\textbf{0.004}\\
\hline
\textbf{12}&\textbf{Rep1}&\textbf{--}&\textbf{0.048}\\
\textbf{12}&\textbf{Rep2}&\textbf{--}&\textbf{0.035}\\
\hline
\end{tabular}
\hspace{0.5cm}
 \begin{tabular}[t]{||c c c c c c c||} 
 \hline
$n$ & $i$ & $s$&$q_{1\rightarrow 0}$ & $q_{0\rightarrow 1}$ & Diff & Emp diff \\[0.5ex]  
\hline\hline 
12&0&0.561&0.046&0.022&0.024&0.027\\
12&1&0.557&0.047&0.020&0.027&0.027\\
12&2&0.530&0.044&0.019&0.025&0.024\\
12&3&0.573&0.047&0.022&0.025&0.024\\
12&4&0.534&0.042&0.018&0.024&0.028\\
12&5&0.574&0.049&0.025&0.024&0.031\\
12&6&0.618&0.054&0.031&0.023&0.018\\
12&7&0.509&0.041&0.016&0.025&0.026\\
12&8&0.541&0.044&0.020&0.024&0.026\\
12&9&0.552&0.043&0.022&0.021&0.019\\
\textbf{12}&\textbf{Avg}&\textbf{0.555}&\textbf{0.046}&\textbf{0.022}&\textbf{0.024}&\textbf{0.025}\\
\textbf{12}&\textbf{Std}&\textbf{0.028}&\textbf{0.004}&\textbf{0.004}&\textbf{0.001}&\textbf{0.004}\\
\hline 
\textbf{12}&\textbf{Rep1}&\textbf{---}&\textbf{0.072}&\textbf{0.053}&\textbf{0.019}&\textbf{---}\\
\textbf{12}&\textbf{Rep2}&\textbf{---}&\textbf{0.050}&\textbf{0.021}&\textbf{0.029}&\textbf{---}\\
\hline
\multicolumn{7}{c}{} \\
 \hline
$n$ & $i$ & $s$&$q_{1\rightarrow 0}$ & $q_{0\rightarrow 1}$ & Diff & Emp diff \\[0.5ex]  
\hline\hline
14&0&0.505&0.041&0.020&0.021&0.021\\
14&1&0.543&0.047&0.024&0.023&0.023\\
14&2&0.520&0.044&0.022&0.022&0.021\\
14&3&0.505&0.040&0.019&0.021&0.021\\
14&4&0.520&0.044&0.022&0.022&0.022\\
14&5&0.465&0.036&0.013&0.023&0.023\\
14&6&0.492&0.041&0.016&0.025&0.023\\
14&7&0.517&0.043&0.020&0.023&0.021\\
14&8&0.495&0.042&0.018&0.024&0.026\\
14&9&0.532&0.045&0.022&0.023&0.023\\
\textbf{14}&\textbf{Avg}&\textbf{0.509}&\textbf{0.042}&\textbf{0.020}&\textbf{0.023}&\textbf{0.022}\\
\textbf{14}&\textbf{Std}&\textbf{0.021}&\textbf{0.003}&\textbf{0.003}&\textbf{0.001}&\textbf{0.001}\\
\hline
\textbf{14}&\textbf{Rep1}&\textbf{---}&\textbf{0.065}&\textbf{0.032}&\textbf{0.033}&\textbf{---}\\
\textbf{14}&\textbf{Rep2}&\textbf{---}&\textbf{0.048}&\textbf{0.022}&\textbf{0.026}&\textbf{---}\\
\hline
\end{tabular}}
\caption{MLE estimations of the parameters $s, q, q_{1\rightarrow 0}$ and $q_{0\rightarrow 1}$ 
for Google's Sycamore with $n=12,14$ qubits.
The column $i$ is the seed of the pseudo-random number generator, each $i$ corresponds to a different circuit.
The left panels show the MLE estimations for the symmetric noise model (Eq.~\eqref{e:symm});
the right panels show the MLE estimations for the asymmetric noise model (Eq.~\eqref{e:asy}).
The lines {\bf Rep1} and {\bf Rep2} provide the reported averaged readout errors based on 2020 Google data file \texttt{som\_params\_by\_qubit.csv} and 2021 Google data file \texttt{readout\_raw\_data.tar}.}

\label {t:googleexp}
\end{table}

\subsection{IBM's Fake Guadalupe simulator}

\begin{table}[H]
\adjustbox{max width=\textwidth}{ \begin{tabular}{||c c c c||} 
 \hline
$n$ & $i$ & $s$ & $q$\\[0.5ex] 
\hline\hline 
12&0&0.988&0.073\\
12&1&0.980&0.065\\
12&2&1.000&0.075\\
12&3&0.989&0.069\\
12&4&0.823&0.056\\
12&5&0.974&0.069\\
12&6&0.934&0.062\\
12&7&0.929&0.063\\
12&8&0.890&0.067\\
12&9&0.938&0.064\\
\textbf{12}&\textbf{Avg}&\textbf{0.945}&\textbf{0.066}\\
\textbf{12}&\textbf{Std}&\textbf{0.052}&\textbf{0.005}\\
\hline 
\textbf{12}&\textbf{Rep}&\textbf{--}&\textbf{0.0218}\\

\hline
\end{tabular}
\hspace{0.5cm}
\begin{tabular}{||c c c c c c c||} 
\hline
$n$ & $i$ & $s$ & $q_{1\rightarrow 0}$ & $q_{0\rightarrow 1}$ & Diff & Emp diff \\[0.5ex]  
\hline\hline 
12&0&0.900&0.084&0.049&0.035&0.092\\
12&1&0.996&0.082&0.050&0.032&0.001\\
12&2&0.999&0.088&0.058&0.030&0.000\\
12&3&0.923&0.081&0.047&0.034&0.129\\
12&4&0.882&0.079&0.043&0.036&-0.021\\
12&5&0.921&0.081&0.050&0.031&0.066\\
12&6&0.911&0.077&0.044&0.033&0.053\\
12&7&0.950&0.079&0.050&0.029&0.018\\
12&8&0.868&0.080&0.047&0.033&0.042\\
12&9&0.926&0.080&0.046&0.034&0.042\\
\textbf{12}&\textbf{Avg}&\textbf{0.928}&\textbf{0.081}&\textbf{0.048}&\textbf{0.033}&\textbf{0.042}\\
\textbf{12}&\textbf{Std}&\textbf{0.041}&\textbf{0.003}&\textbf{0.004}&\textbf{0.002}&\textbf{0.043}\\
\hline 
\textbf{12}&\textbf{Rep}&\textbf{--}&\textbf{0.028}&\textbf{0.007}&\textbf{0.021}&\textbf{--}\\
\hline
\end{tabular}}
\caption{MLE estimations of the (effective) readout error rates for IBM's Fake Guadalupe simulator with $n=12$ qubits.
The column $i$ is the seed of the pseudo-random number generator, each $i$ corresponds to a different circuit.
The left panel shows the MLE estimations for the symmetric noise model (Eq.~\eqref{e:symm});
the right panel shows the MLE estimations for the asymmetric noise model (Eq.~\eqref{e:asy}).
Here, the MLE estimations for effective readout error rates appear to be considerably larger than the physical readout errors.
The average value of readout error (for the 12 qubits used in our circuits) are $q_{1\rightarrow 0}=0.0278$, $q_{0\rightarrow 1}=0.0072$. 
The MLE estimations are on average $q_{1\rightarrow 0}=0.081$, $q_{0\rightarrow 1}=0.048$.
Both values are considerably higher than the physical readout errors (reported in the last line).
The average of these two quantities is quite close to the MLE estimation for $q=0.066$ in the 2-parameter model.\\
The average empirical difference between ratios of 1's and 0's is 0.042, the individual values are rather volatile.
This value is considerably larger than the difference between the two MLE-estimated parameters which is 0.033 (for those, the individual values are rather stable).
These values are considerably larger than the average difference between the reported readout error rates, which is 0.0207.)
}
\label {table:fake_s_q}
\end{table}

\subsection{Google's QVM Weber simulator}
\enlargethispage{1cm}

\begin{table}[H]
\adjustbox{max width=.93\textwidth}{ \centering
 \begin{tabular}[t]{||c c c c||} 
 \hline
$n$ & $i$ & $\phi_g$ & $q$\\[0.5ex] 
\hline\hline 
12&0&0.778&0.069\\
12&1&0.629&0.054\\
12&2&0.606&0.050\\
12&3&0.525&0.041\\
12&4&0.709&0.063\\
12&5&0.782&0.070\\
12&6&0.496&0.036\\
12&7&0.586&0.049\\
12&8&0.599&0.050\\
12&9&0.560&0.045\\
\textbf{12}&\textbf{Avg}&\textbf{0.627}&\textbf{0.053}\\
\textbf{12}&\textbf{Std}&\textbf{0.094}&\textbf{0.011}\\
\hline
\textbf{12}&\textbf{Rep0}&\textbf{--}&\textbf{0.038}\\
\textbf{12}&\textbf{Rep1}&\textbf{--}&\textbf{0.041}\\
\textbf{12}&\textbf{Rep2}&\textbf{--}&\textbf{0.058}\\
\hline
\multicolumn{4}{c}{} \\
\hline
$n$ & $i$ & $\phi_g$ & $q$\\[0.5ex] 
\hline\hline 
14&0&0.512&0.045\\
14&1&0.542&0.049\\
14&2&0.471&0.040\\
14&3&0.606&0.056\\
14&4&0.576&0.053\\
14&5&0.518&0.046\\
14&6&0.533&0.049\\
14&7&0.419&0.033\\
14&8&0.630&0.058\\
14&9&0.511&0.046\\
\textbf{14}&\textbf{Avg}&\textbf{0.532}&\textbf{0.047}\\
\textbf{14}&\textbf{Std}&\textbf{0.059}&\textbf{0.007}\\
\hline
\textbf{14}&\textbf{Rep0}&\textbf{--}&\textbf{0.038}\\
\textbf{14}&\textbf{Rep1}&\textbf{--}&\textbf{0.039}\\
\textbf{14}&\textbf{Rep2}&\textbf{--}&\textbf{0.055}\\
\hline
\end{tabular}
\hspace{0.5cm}
 \begin{tabular}[t]{||c c c c c c c||} 
 \hline
$n$ & $i$ & $\phi_g$ & $q_{1\rightarrow 0}$ & $q_{0\rightarrow 1}$ &Diff & Emp diff \\[0.5ex]  
\hline\hline 
12&0&0.596&0.078&0.021&0.057&0.061\\
12&1&0.641&0.084&0.027&0.057&0.056\\
12&2&0.635&0.082&0.025&0.057&0.056\\
12&3&0.549&0.073&0.016&0.057&0.057\\
12&4&0.582&0.077&0.020&0.057&0.061\\
12&5&0.553&0.072&0.016&0.056&0.063\\
12&6&0.625&0.082&0.025&0.057&0.053\\
12&7&0.574&0.076&0.019&0.057&0.057\\
12&8&0.541&0.071&0.013&0.058&0.060\\
12&9&0.594&0.078&0.021&0.057&0.056\\
\textbf{12}&\textbf{Avg}&\textbf{0.589}&\textbf{0.077}&\textbf{0.020}&\textbf{0.057}&\textbf{0.058}\\
\textbf{12}&\textbf{Std}&\textbf{0.034}&\textbf{0.004}&\textbf{0.004}&\textbf{0.000}&\textbf{0.003}\\ 
\hline
\textbf{12}&\textbf{Rep0}&\textbf{--}&\textbf{0.055}&\textbf{0.023}&\textbf{0.057}&\textbf{--}\\
\textbf{12}&\textbf{Rep1}&\textbf{--}&\textbf{0.069}&\textbf{0.012}&\textbf{0.057}&\textbf{--}\\
\textbf{12}&\textbf{Rep2}&\textbf{--}&\textbf{0.100}&\textbf{0.015}&\textbf{0.085}&\textbf{--}\\
\hline
\multicolumn{7}{c}{} \\
\hline
$n$ & $i$ & $\phi_g$ & $q_{1\rightarrow 0}$ & $q_{0\rightarrow 1}$ & Diff & Emp diff \\[0.5ex]  
\hline\hline
14&0&0.502&0.071&0.016&0.055&0.056\\
14&1&0.492&0.070&0.014&0.056&0.057\\
14&2&0.513&0.073&0.018&0.055&0.054\\
14&3&0.622&0.086&0.030&0.056&0.055\\
14&4&0.557&0.078&0.023&0.055&0.056\\
14&5&0.541&0.077&0.021&0.056&0.056\\
14&6&0.558&0.079&0.024&0.055&0.055\\
14&7&0.511&0.073&0.017&0.056&0.054\\
14&8&0.540&0.076&0.020&0.056&0.058\\
14&9&0.507&0.073&0.017&0.056&0.057\\
\textbf{14}&\textbf{Avg}&\textbf{0.534}&\textbf{0.076}&\textbf{0.020}&\textbf{0.056}&\textbf{0.056}\\ 
\textbf{14}&\textbf{Std}&\textbf{0.037}&\textbf{0.004}&\textbf{0.004}&\textbf{0.000}&\textbf{0.001}\\
\hline 
\textbf{14}&\textbf{Rep0}&\textbf{--}&\textbf{0.055}&\textbf{0.023}&\textbf{0.032}&\textbf{--}\\
\textbf{14}&\textbf{Rep1}&\textbf{---}&\textbf{0.067}&\textbf{0.011}&\textbf{0.056}&\textbf{---}\\
\textbf{14}&\textbf{Rep2}&\textbf{---}&\textbf{0.095}&\textbf{0.014}&\textbf{0.081}&\textbf{---}\\
\hline
\end{tabular}}
\caption{MLE estimations of the (effective) readout error rates for Google's QVM Weber simulator with $n=12,14$ qubits.
The column $i$ is the seed of the pseudo-random number generator, each $i$ corresponds to a different circuit.
The left panels show the MLE estimations for the symmetric noise model (Eq.~\eqref{e:symm});
the right panels show the MLE estimations for the asymmetric noise model (Eq.~\eqref{e:asy}).
In the table, Rep0 corresponds to the averaged readout error rates reported in the supplement of Google's paper \cite {Aru+19},  Rep1 corresponds to the readout errors in the simulator's documentation under the category ``alone" and Rep2 corresponds to those errors under the category ``parallel".   
}
\label {t:weber}
\end{table}

\newpage

\begin{samepage}

\subsection{5-qubit circuits on IBM quantum processors}
\enlargethispage{2cm}

\subsubsection*{Nairobi}
\adjustbox{max width=\textwidth}{ \begin{tabular}{||c c c c||} 
 \hline
$n$ & $i$ & $\phi_g$ & $q$\\[0.5ex] 
\hline \hline 
5&0&1.000&0.083\\
5&1&0.635&0.000\\
5&2&0.732&0.024\\
5&3&1.000&0.089\\
5&4&0.945&0.053\\
5&5&0.721&0.033\\
5&6&1.000&0.060\\
5&7&1.000&0.078\\
5&8&0.768&0.018\\
5&9&0.731&0.001\\
\textbf{5}&\textbf{Avg}&\textbf{0.853}&\textbf{0.044}\\
\textbf{5}&\textbf{Std}&\textbf{0.140}&\textbf{0.032}\\
\hline 
\end{tabular}
\hspace{0.5cm}
 \begin{tabular}{||c c c c c c c||} 
 \hline
$n$ & $i$ & $\phi_g$ & $q_{1\rightarrow 0}$ & $q_{0\rightarrow 1}$ & Diff & Emp diff \\[0.5ex]  
\hline \hline 
5&0&0.999&0.118&0.056&0.062&0.103\\
5&1&0.732&0.046&0.000&0.046&0.045\\
5&2&0.732&0.023&0.025&-0.002&0.024\\
5&3&0.953&0.103&0.058&0.045&0.069\\
5&4&0.852&0.070&0.012&0.058&0.151\\
5&5&0.959&0.120&0.042&0.078&-0.008\\
5&6&0.984&0.086&0.030&0.056&0.084\\
5&7&0.999&0.109&0.046&0.063&0.058\\
5&8&0.779&0.055&0.000&0.055&0.065\\
5&9&0.790&0.036&0.000&0.036&0.016\\
\textbf{5}&\textbf{Avg}&\textbf{0.878}&\textbf{0.077}&\textbf{0.027}&\textbf{0.050}&\textbf{0.061}\\
\textbf{5}&\textbf{Std}&\textbf{0.107}&\textbf{0.034}&\textbf{0.022}&\textbf{0.020}&\textbf{0.045}\\
\hline 
\end{tabular}}

\subsubsection*{Jakarta}

\begin{table}[H]
\adjustbox{max width=\textwidth}{\begin{tabular}{||c c c c||} 
\hline
$n$ & $i$ & $\phi_g$ & $q$\\[0.5ex] 
\hline\hline 
5&0&0.851&0.067\\
5&1&1.000&0.095\\
5&2&0.896&0.088\\
5&3&0.793&0.063\\
5&4&0.685&0.058\\
5&5&0.946&0.109\\
5&6&0.855&0.073\\
5&7&0.550&0.021\\
5&8&0.566&0.012\\
5&9&0.867&0.083\\
\textbf{5}&\textbf{Avg}&\textbf{0.801}&\textbf{0.067}\\
\textbf{5}&\textbf{Std}&\textbf{0.145}&\textbf{0.029}\\
\hline 
\end{tabular}
\hspace{.5cm}
\begin{tabular}{||c c c c c c c||} 
\hline
$n$ & $i$ & $\phi_g$ & $q_{1\rightarrow 0}$ & $q_{0\rightarrow 1}$ & Diff & Emp diff \\[0.5ex]  
\hline\hline 
5&0&0.758&0.099&0.013&0.086&0.124\\
5&1&0.943&0.114&0.067&0.047&0.079\\
5&2&0.870&0.094&0.073&0.021&0.052\\
5&3&0.782&0.063&0.060&0.003&0.079\\
5&4&0.569&0.059&0.007&0.052&0.079\\
5&5&0.848&0.117&0.076&0.041&0.123\\
5&6&0.745&0.077&0.037&0.040&0.102\\
5&7&0.545&0.039&0.006&0.033&0.034\\
5&8&0.569&0.021&0.007&0.014&0.021\\
5&9&0.852&0.071&0.095&-0.024&-0.040\\
\textbf{5}&\textbf{Avg}&\textbf{0.748}&\textbf{0.075}&\textbf{0.044}&\textbf{0.031}&\textbf{0.065}\\
\textbf{5}&\textbf{Std}&\textbf{0.134}&\textbf{0.030}&\textbf{0.032}&\textbf{0.028}&\textbf{0.050}\\
\hline 
\end{tabular}}
\caption{MLE estimations of the (effective) readout error rates for IBM's 7-qubit Nairobi and Jakarta quantum processors with $n=5$-qubit circuits.
The column $i$ is the seed of the pseudo-random number generator, each $i$ corresponds to a different circuit.
The left panels show the MLE estimations for the symmetric noise model (Eq.~\eqref{e:symm});
the right panels show the MLE estimations for the asymmetric noise model (Eq.~\eqref{e:asy}).
Here, the 5-qubit averages $q_{1 \rightarrow 0}$ are 0.0373 and 0.0360,
and the 5-qubit averages $q_{0 \rightarrow 1}$ are 0.0138 and 0.0135.}
\label {t:ibm_q_nj}
\end{table}

\end{samepage}

\section {Extending the Fourier Techniques to Asymmetric Readout Errors}
\label {s:fourier}

Let
$$\Omega_n = \{x=(x_1,x_2,\dots,x_n) : x_i \in \{0,1\}, i=1,2,\dots,n\}.$$
$\Omega_n$ is referred to as the discrete $n$-dimensional cube.
(Elements in $\Omega_n$ are bitstrings of length $n$ of zeroes and ones.)
In \cite {KRS24} we considered real functions on $\Omega_n$ and their Fourier--Walsh expansion.
The Fourier--Walsh representation gave us a useful representation of the symmetric noise model \eqref{e:symm}, a refinement of the XEB fidelity estimator, and FFT-based methods for computing our estimators from Section 6 of \cite {RSK22}. 
Here, we briefly describe how to extend the Fourier method for asymmetric errors.

For a real number $p$, $0 \le p \le 1$, let $\mu_p$ be the $p$-biased measure on $\Omega_n$, namely, 
\begin {equation} 
\mu_p(x_1,x_2,\dots,x_n)=p^k(1-p)^{n-k},
\end {equation}
where $k=x_1+x_2+ \cdots +x_n$.

\subsection {$p$-biased Fourier--Walsh basis}

Let $p,\, 0<p<1$ be a real number. 
The space $L_2(\mu_p)$ is the space of real functions on $\Omega_n$ with the inner product 
$$\langle f,g\rangle =\mathbb E _{x \sim \mu_p} f(x)g(x).$$
For $S \subset [n]$ we let
$$\chi_S^p=\prod_{i \in S} \frac {x_i-p}{\sqrt {p(1-p)}}.$$
These functions form an orthonormal basis for $L_2(\mu_p)$.
For a real function $f \in L_2(\mu_p)$ we write its expansion
$$ f=\sum _{S \subset [n]} \widehat f(S) \chi_S^p$$
in terms of this basis.

\subsection{One-sided noise operator}

The one-sided noise operator $\mathrm{T}^{p\to q}$ is defined on $\Omega_n$ as follows.
Given $x \in \Omega_n$, $N(x)$ is a distribution on $\Omega_n$ where every coordinate with value 0 stays the same
and every coordinate with value 1 becomes a 0 with probability $\frac {q-p}{q}$, independently over all $i$. 
$$\mathrm{T}^{p\to q}(f)(x)= \mathbb E _{y\sim N(x)}f(y).$$ 

{\bf Remarks:} 

\begin {enumerate}
\item If $x \sim \mu_q$ and $y \sim N(x)$ then the marginal distribution for $y$ is $\mu_p$. 

\item If $f$ is monotone then
$$\langle \mathrm{T}^{p\to q},f\rangle_{\mu(q)}=\mathbb E_{x \sim \mu_q, y \sim N(x)}f(x)f(y) =\mu_p(f).$$
Our argument will be based on finding conditions implying that $f(x)$ and $f(y)$ are nearly independent. 

\end {enumerate}

The following simple Lemma is taken from \cite {Lif20}.
\begin{lem}
\label{lem:Eigenvalues of one sided noise operator}
Let $p<q\in\left(0,1\right)$,
$\rho=\sqrt{\frac{p\left(1-q\right)}{q\left(1-p\right)}}.$
Let $f\in L_{2}\left(\left\{ 0,1\right\} ^{n},\mu_{p}\right)$ and write 
\[
f=\sum\hat{f}\left(S\right)\chi_{S}^{p}.
\]
Then 
\[
\mathrm{T}^{p\to q}f=\sum\rho^{\left|S\right|}\hat{f}\left(S\right)\chi_{S}^{q}
\]
\end{lem}

\subsection {Asymmetric noise model}

Starting with a probability distribution ${\cal P}$ consider an asymmetric noise model
$R(q_{1\rightarrow 0},q_{0\rightarrow 1})({\cal P}),$ and suppose that $q_{0\rightarrow 1}<q_{1\rightarrow 0}.$
We can consider this operation as a composition of a symmetric noise $R(q_{0\rightarrow 1},q_{0\rightarrow 1})$ with a one-sided noise $\mathrm{T}^{\frac{1}{2}\to q}$ for an appropriately chosen value of $q$.
This gives us a presentation of the noisy distributions in terms of the $p$-biased Fourier--Walsh basis of $L_2(\mu_p)$ where $p=1/2+(q_{0\rightarrow 1}-q_{1\rightarrow 0})/2$.

{\bf Remark:} The FFT algorithm relies only on the tensor structure of the Fourier--Walsh transforms and it therefore extends also to the $p$-biased case and to the case where the readout error rates vary over different qubits.
(We thank Elchanan Mossel for helpful comments on this point.)

\section{Temporal variation of the proportion of measured 1's during the measurement of bitstrings}
\label{s:temporal_instability}

The figures in this section show the temporal instability of the proportion of measured 1's during the course of an experiment.
We observe such instabilities for Sycamore-53, -67, and -69 as well as for Zuchongzhi-56.
Since Google's initialization/measurement data do not display a pronounced temporal variation (see Fig.~\ref{fig_q10_q01_SYC-53_by_hundredth.pdf}), the temporal instability in the proportion of measured 1's does not seem to originate from a temporal variation of preparation or readout errors, but from an instability of applied gates during the cycles.

\begin{figure*}[!ht]
    \vspace{-1cm}
    \centering
    \hspace*{-2cm}
    \includegraphics[width=7.1in]{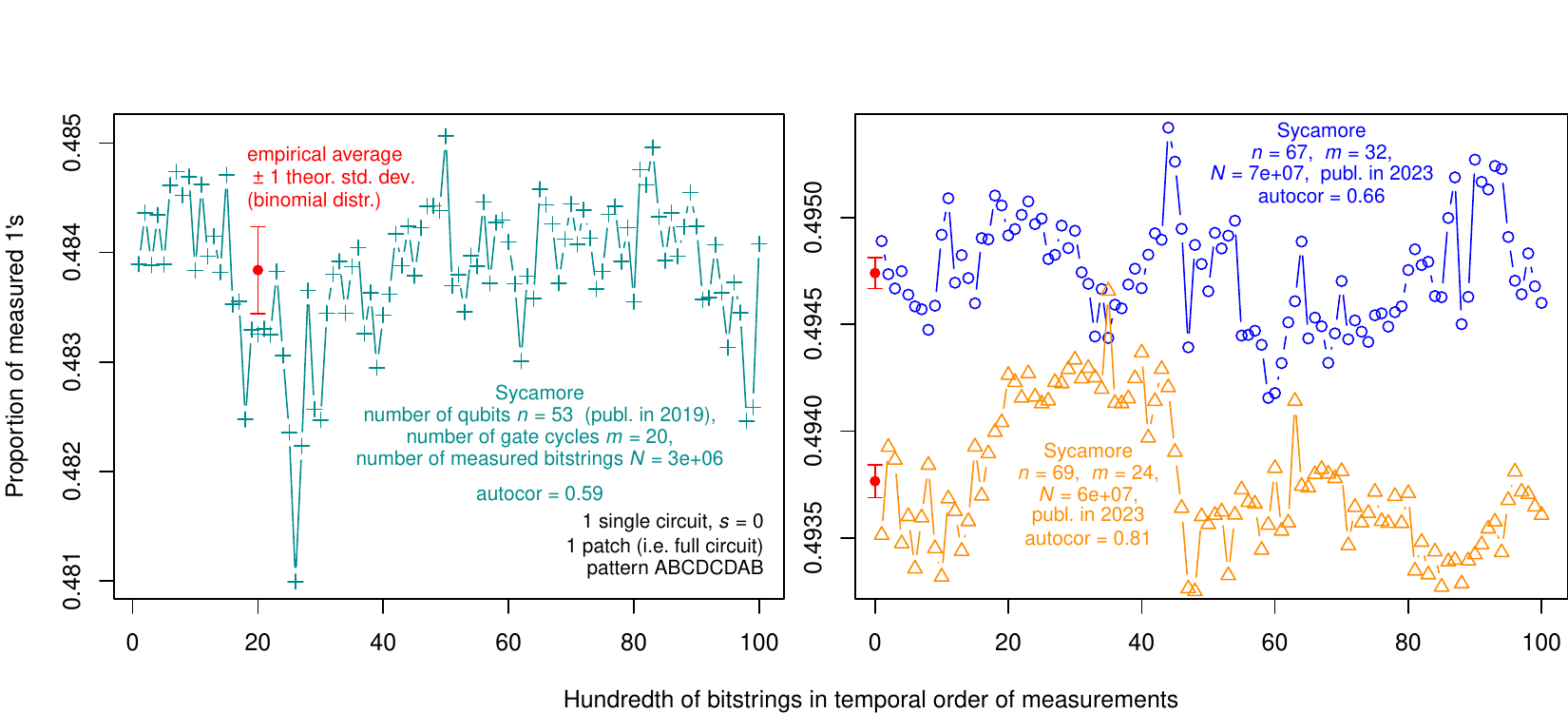}
    \caption{\small {\textbf{Temporal instability in the proportion of measured 1's during the measurement of bitstrings for three Sycamore circuits.}
    The left panel shows the proportion of 1's for number of qubits $n=53$ \cite {Aru+19}, the right panel for number of qubits $n=67$ and $n=69$ (both published in 2023).
    The hundredth $i$ on the horizontal axis means that the corresponding proportion of 1's is for those of the $N$ measured bitstrings with index from $((i-1) \times N/100 + 1)$ to $i \times N/100$, where the bitstrings are indexed in temporal order of their measurement.
    So $i=1$ contains the first hundredth of the measured bitstrings, whereas $i=100$ contains the last hundredth of the measured bitstrings.
    For each of the three shown values of $n$, the bitstrings are measured for a single circuit.
    All three circuits are 1-patch circuits (i.e. full circuits) with the non-simplifiable coupler activation pattern ABCDCDAB.
    The stated (normalized) autocorrelation is the correlation between the two vectors proportion(hundredth $i$) and proportion(hundredth $i+1$) for $i \in \{1, \dots, 99\}$.
    For short time scales comprising the measurement of only a few hundredths of bitstrings, the standard deviation of measured 1's for each of the three circuits agrees with the standard deviation of a binomial distribution (red error bars for each circuit).
    But globally, on larger time scales comprising the measurement of many hundredths, the mean proportion of 1's shows an unstable behavior, shifting upwards and downwards by many theoretical standard deviations.
    Ideally, the proportion of measured 1's is expected to be distributed around a constant average, where each value is independent from its predecessor value.
    This would result in an autocorrelation near zero.
    In contrast to this ideal, the relatively large positive actual autocorrelations are an indication of unstable readout errors or of unstable other experimental parameters over time during the initial state preparation, during the gate cycles or during the bitstring measurements.}}
    \label{fig_proportion_ones_SYC_by_hundredth}
\end{figure*}

\begin{figure*}[!ht]
    \centering
    \includegraphics[width=4.2in]{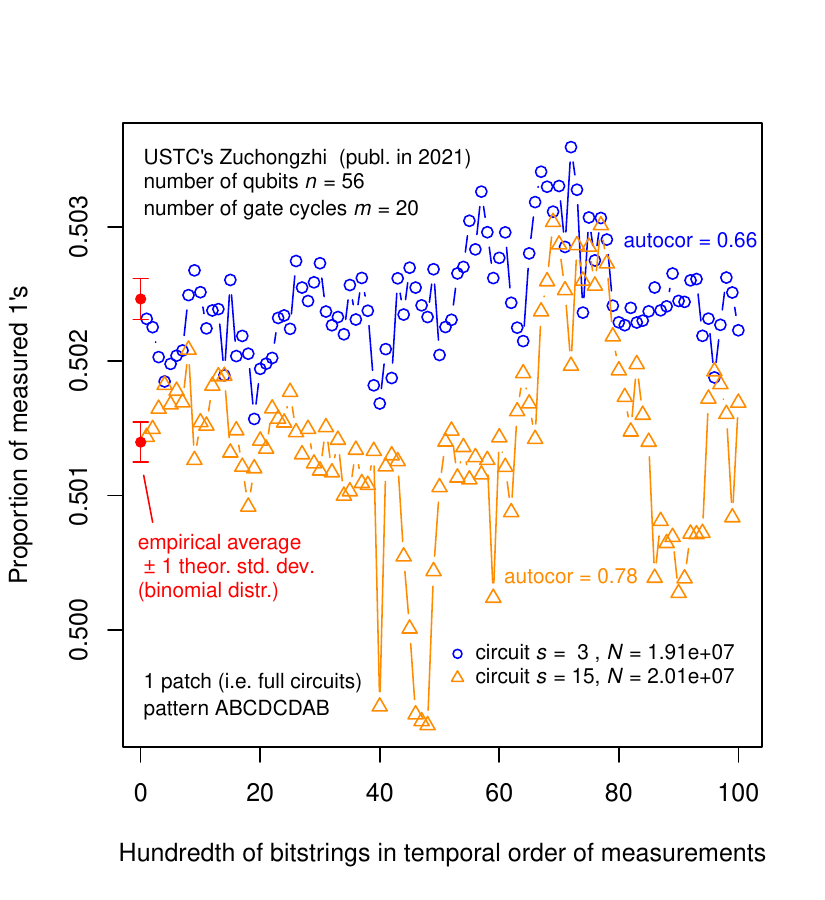}
    \caption{\textbf{Temporal instability in the proportion of measured 1's during the measurement of bitstrings for two of USTC's Zuchongzhi-56 circuits.}
    In contrast to Google's Sycamore, most of Zuchongzhi's proportion of 1's are greater than 0.5.
    The other characteristics of Zuchongzhi's temporal behavior of its proportion of 1's are similar to Sycamore:
    A "local" agreement of variation with a binomial distribution, but a "global" unstable behavior with shifts over many theoretical standard deviations.
    This leads to a large positive autocorrelation, also similar to Sycamore.}
    \label{fig_proportion_ones_ZCZ-56_by_hundredth}
\end{figure*}

\begin{figure*}[!ht]
    \centering
    \hspace*{-2cm}
    \includegraphics[width=7.1in]{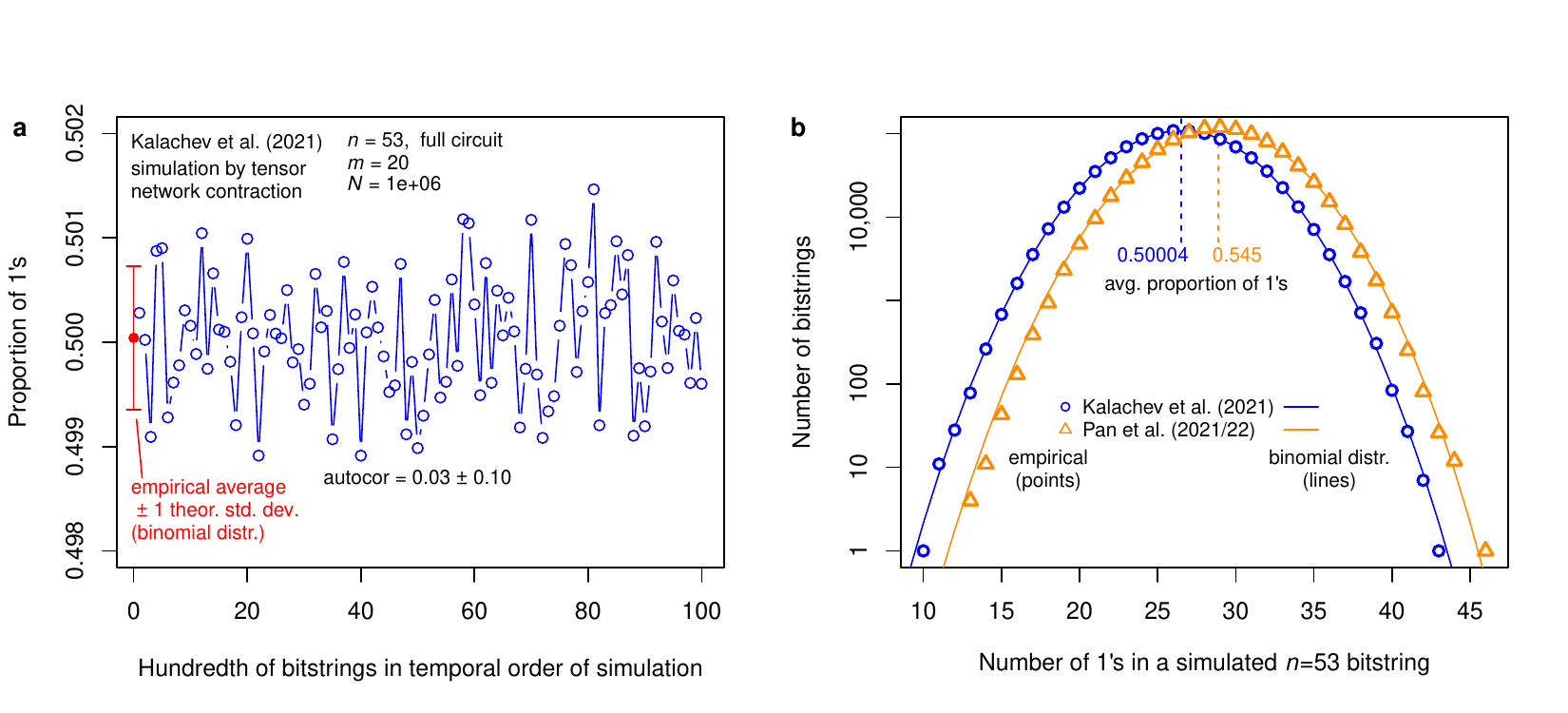}
    \caption{\textbf{Proportion of 1's in simulation of bitstrings by Kalachev et al.'s ("Classical sampling ...") and Pan et al.'s ("Solving the sampling problem ...") tensor network contraction for $n=53$ qubits. a,}
    Proportion of 1's for hundredths of all simulated bitstrings by Kalachev et al. in temporal order of simulation.
    As expected, the simulated proportion of 1's shows no temporal instability and thus a vanishing autocorrelation, contrasting the experimental Sycamore and Zuchongzhi data.
    The variation is in agreement with a binomial distribution of the number of measured 1's.
    \textbf{b,}
    Distribution of the number of 1's in the bitstrings simulated by Kalachev et al. ($m=20$ cycles) and Pan et al.
    For Pan et al., the average proportion of 1's is very large at 0.545.
    For both publications, the number of 1's agrees well with a binomial distribution.}
    \label{fig_proportion_ones_Kalachev_Pan}
\end{figure*}

\begin{figure*}[!ht]
    \centering
    \hspace*{-2cm}
    \includegraphics[width=7.1in]{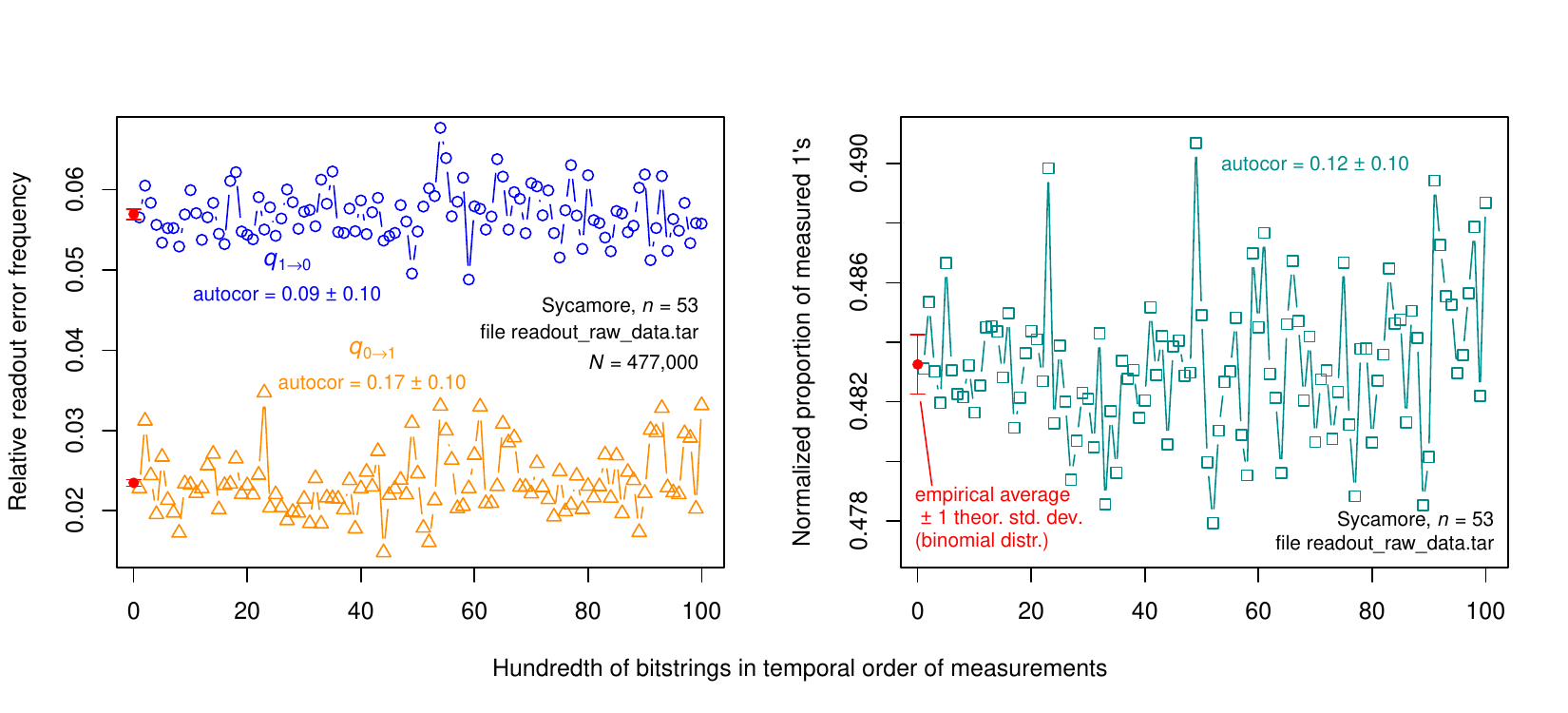}
    \caption{\textbf{Asymmetric readout error relative frequencies (left panel) and resulting normalized proportion of measured 1's (right panel) for Google's Sycamore $n=53$ qubit initialization/measurement data \texttt{readout\_raw\_data.tar}.}
    The hundredth $i \in \{1, \dots, 100\}$ on the horizontal axis means that the corresponding proportion of 1's is for those of the $N$ measured bitstrings with index from $((i-1) \times N/100 + 1)$ to $i \times N/100$, where the bitstrings are indexed in temporal order of their measurement.
    The variation of the relative readout errors is much larger than the theoretically expected standard deviation of a binomial distribution (red error bars on the left of each panel).
    This can be explained by a larger asymmetric readout error variation of individual qubits.
    The relative asymmetric readout error frequencies and also the resulting proportion of 1's display no pronounced time-dependent behavior, which also shows in the near-zero autocorrelations.
    The normalized proportion of 1's was estimated from the asymmetric readout error relative frequencies as $1/2 + (q_{0\rightarrow1}-q_{1\rightarrow0})/2$.
    This normalization is necessary, since the measured "output" bitstrings are very similar to the initialized "input" state bitstrings and these input bitstrings were not selected by Google to have a mean proportion of 1's of exactly 1/2.
    The pronounced temporal instability in the proportion of measured 1's of Sycamore and Zuchongzhi bitstrings shown in Figs.~\ref{fig_proportion_ones_SYC_by_hundredth} and \ref{fig_proportion_ones_ZCZ-56_by_hundredth} thus does not seem to primarily originate from a temporal variation of preparation or readout errors, but from an instability of applied gates during the cycles.}
    \label{fig_q10_q01_SYC-53_by_hundredth.pdf}
\end{figure*}

\bigskip

{\small
\noindent
Gil Kalai,  Hebrew University of Jerusalem, Einstein Institute of Mathematics, and\\  Reichman University, Efi Arazi School of Computer Science. \\ {\tt gil.kalai@gmail.com}.

\medskip

\noindent
Tomer Shoham,  Hebrew University of Jerusalem, Federmann Center for the Study of Rationality and Department of Computer Science.\\ {\tt tomer.shohamm@gmail.com}.

\medskip

\noindent
Carsten Voelkmann, Munich University of Applied Sciences, Department of Engineering and Management.\\ {\tt carsten.voelkmann@hm.edu}.
}


\clearpage

\begin{thebibliography}{99}

\bibitem{AGLLV22}
D. Aharonov, X. Gao, Z. Landau, Y. Liu, U. Vazirani, 
A polynomial-time classical algorithm for noisy random circuit sampling, arXiv:2211.03999 (2022). 

\bibitem {Aru+19}
F. Arute et al.,
Quantum supremacy using a programmable superconducting processor, {\it Nature} 574 (2019), 505--510.


\bibitem {Blu+23}
D. Bluvstein, S. J. Evered, A. A. Geim, et al.,
Logical quantum processor based on reconfigurable atom arrays, {\it Nature} 626 (2024), 58--65. arXiv:2312.03982.





\bibitem{FGG+23}  
B. Fefferman, S. Ghosh, M. Gullans, K. Kuroiwa, K. Sharma,
Effect of non-unital noise on random circuit sampling, arXiv:2306.16659 (2023).


\bibitem{Gao+21}
X. Gao, M. Kalinowski, C.-N. Chou, M. D. Lukin, B. Barak, and S. Choi, 
Limitations of linear cross-entropy as a measure for quantum advantage, arXiv:2112.01657 (2021).





\bibitem {Kal24}
G. Kalai,
The argument against quantum computers, the quantum laws of nature, and Google’s supremacy claims, in: {\it The Intercontinental Academia Laws: Rigidity and Dynamics} (M. J. Hannon and E. Z. Rabinovici (eds.)), 
World Scientific 2024. arXiv:2008.05188.


\bibitem{KalKin14}
G. Kalai and G. Kindler,
Gaussian noise sensitivity and BosonSampling, arXiv:1409.3093 (2014).

\bibitem {KRS22d}
G. Kalai, Y. Rinott, and T. Shoham,
Google’s quantum supremacy claims: data, documentation, and discussion, arXiv:2210.12753 (2022). 

\bibitem {KRS23}
G. Kalai, Y. Rinott, and T. Shoham,
Questions and concerns about Google’s quantum supremacy claim, arXiv:2305.01064 (2023). 

\bibitem {KRS24}
G. Kalai and Y. Rinott, and T. Shoham,
Random Circuit Sampling: Fourier--Walsh Expansion and Statistics, arXiv:2404.00935 (2024).

\bibitem {KSV25} G. Kalai, T. Shoham, and C. Voelkmann,
Further Statistical Study of NISQ Experiments, arXiv:2512.10722 (2025).


\bibitem {Lif20}
N. Lifshitz,
Hypergraph removal lemmas via robust sharp threshold theorems, Discrete Analysis 2020:10, 46 pp, arXiv:1804.00328.
 




\bibitem {RSK22}
Y. Rinott, T. Shoham, and G. Kalai,
Statistical aspects of the quantum supremacy demonstration, {\it Statistical Science} 37 (2022), 322--347. arXiv version: arXiv:2008.05177 (2020).

\end{thebibliography}
\end{document}